\documentclass{article}

\usepackage{PRIMEarxiv}
\usepackage[utf8]{inputenc} 
\usepackage[T1]{fontenc}    
\usepackage{charter}
\usepackage{hyperref}       
\usepackage{url}            
\usepackage{booktabs}       
\usepackage{amsfonts}       
\usepackage{nicefrac}       
\usepackage{microtype}      
\usepackage{lipsum}
\usepackage{fancyhdr}       
\usepackage{graphicx}       
\graphicspath{{media/}}     
\title{Exbodiment: The Mind Made Matter}

\author{
  David C. Krakauer \\
  Santa Fe Institute\\
  1399 Hyde Park Road \\
  Santa Fe, NM, USA, 87501\\
  \texttt{dk@santafe.edu} \\
  \\
}

\begin{document}
\maketitle

\begin{abstract}
Exbodiment describes mind outsourced to engineered matter and how  matter reeducates mind \footnote{This manuscript is an extended version of the article Mind Made Matter that appeared in Volume 1 of the 2022 journal Return.}. The constraints of exbodied matter encode elements of thought, channel decision-making, and constitute an important part of an extended computational phenotype. Here I provide an introduction and brief cultural history of exbodiment in music, natural history, cognition, and astrobiology. The "Helix of Exbodiment" is introduced to illustrate continuous feedback between mind and matter. 
\end{abstract}

\begin{quote}
    "In spite of its apparent seduction, a metaphysics of the body is as reductionist as one grounded on the electron" \cite{Smith1998-zt}
\end{quote}

\section{Musico-material Constraint}

\subsection{Chopin's Preludes}

Chopin’s Preludes for solo piano are regarded as some of the more trailblazing compositions in the history of piano music.
The preludes were written between 1835 and 1839 in a remote and chilly Carthusian cell in the monastery of Valldemosa in Palma, Majorca. Chopin is reported to have described his cell as “shaped like a tall coffin”. \cite{Rowland2020-ob}

Chopin's Preludes followed the template of Bach’s 48 Preludes and Fugues published in his Well-Tempered Clavier from 1722. The novelist George Sand, Chopin’s partner at the time wrote, "His genius was filled with the mysterious sounds of nature, but transformed into sublime equivalents in musical thought…” \cite{Sand1991-jy} As it turns out – both Natural history and instrumental history.

No workroom could be worse equipped for acoustics than the cell in Valldemosa. And the piano on which Chopin composed the Preludes was locally constructed by an otherwise unknown carpenter, Jaun Bauza. The Piano was by the standards of its time a dud. According to Paul Kildea, the author of a thoughtful history of Chopin’s Bauza, “it was out of date before it was completed… unable to support thicker or longer strings, greater tension or larger compass..its wooden frame hostage to the island’s fierce climate”. \cite{Kildea2018-mu}

It was on this wretched instrument that Chopin wrote the Preludes and not on the far more sonorous Pleyel pianos on which he was accustomed to composing. It is as if the further Chopin moved from his Platonic musical ideals, into the texture and tension of physicality, the more his imagination found the space to become creative. In Malta he composed not as Bach’s epigone but as his elemental heir apparent.

\subsection{Jarret's Koln Concert}

From the pub-like keys of a practice piano in a lugubrious Koln winter in 1974 Keith Jarrett improvised a composition that would go on to become the best-selling Jazz album of all time. \cite{Elsdon2013-hg} In reflecting on the condition of his piano, Jarrett described it as not worthy of recording.
It was over a hundred and thirty years after Chopin that the thirty-year-old Jarrett planed to perform a solo piano recital on a Bosendorfer 290 Imperial concert grand. On the night of the performance, Jarret was delivered a vastly less edifying, and out of tune to boot, Bosendorfer baby grand. As Jarret describes it, "we had the wrong piano rented…I had a Bosendorfer I really didn't like..I had not slept for two days..everything was wrong. When you have problems one after another you forget what you are doing" (From : https://www.bbc.co.uk/programmes/m0011f4y)

\subsection{In Praise of Imperfection}

Like Chopin's Bauza, Jarret's physical instrument had "ideas of its own". And these did not align with anyone's ideal tonalities. The mini grand sounds less like a piano and more like a fortepiano and it "wants" to be played like a harpsichord. Jarrett achieved his effects throughout the first movement by abstaining from the use of the sustain pedal. Jarrett’s imposed isorhythms of involuntary serialism are reminiscent of the contemporary experiments of Glass, Riley, and Reich with arpeggios and polyrhythms.

It took a harmonically compromised piano to lure Jarrett away from experimental jazz into a space hitherto dominated by modern classical composers. And it took a similar experience to dislocate Chopin into the baroque lyric of his Preludes. Both Chopin and Jarrett were experiencing \emph{exbodiment} - that state of cognition characterized by a collaboration between organismal mind and environmentally manipulated matter. One where matter becomes a computational collaborator and not merely a mechanical drudge.

\section{The Hymenopteran Sextant} 

In 1973 the Nobel prize for physiology or medicine went to a rather unlikely triumvirate: Karl von Frisch, Konrad Lorenz, and Niko Tinbergen. This was the first and the last time that the prize was awarded for animal behaviour – ethology. To put this in perspective, all Nobel prizes in the previous several decades had been awarded for reductionist approaches to molecular medicine. Following the 1973 award not a single prize has been awarded to whole organism biology. For the Nobel committee, Physiology and Medicine seem to be synonyms for cell biology.

Even more extraordinary is that Karl von Frisch had at no point in his career shown any interest whatsoever in human beings or their cells. The closest he got to cells were the hexagonal lattices of honeybee hives. This contrasts favourably with Konrad Lorenz who had a rather unhealthy fascination with human copulation, aggression, and domestication, which would ultimately draw him into the delirious orbit of Nazi science \cite{Weikart2013-bv}. Niko Tinbergen who fought for the resistance was imprisoned during the war.

\subsection{Ideal Tools}

In a 1968 review of the life and work of von Frisch, the late Edward Wilson wrote, "Every successful scientist has a small number of personal tools with which he levers discoveries out of nature. Von Frisch had two in which he attained great mastery. The first was the repeated exploitation of the passage of honeybees from nest to flowers and back again, a complex sequence of behavioural events that is nonetheless easy to manipulate and to monitor. The second was the method of Pavlovian training, by which von Frisch associated the stimuli to be studied with a subsequent reward of food." \cite{Wilson1968-xv}. 

And using these experimental tools von Frisch was able to discover and decipher one of Nature's most extraordinary signaling systems, the so-called dance language of the honeybee. In brief, bees returning from a source of pollen communicate the direction and the distance of nutritive pollen to the colony through an elaborate sequence of movements set against a honeycomb. The angular deviation of a forward waggle motion from the overhead position of the sun in relation to the comb describes the orientation of a food source. The amplitude of the waggle encodes the distance. Thus, the stable position of the comb allows the bee to use gravity as a reference. \cite{Von_Frisch1956-te}. 

\subsection{Natural Historical Instruments}

In many ways the hive with its combs serves as an inclinometer equiped astrolabe \cite{von-Frisch1974-pj}. The astrolabe is a simple two-dimensional model of the celestial sphere which can be used to measure the precise position of celestial objects above the horizon. And in this way determine time and latitude. In order that an astrolabe function reliably, a plumb line is required to ensure that it is positioned perpendicular to the earth or ocean. The hive combs likewise need to preserve a fixed angle with respect to gravity, whereby the orientation of the waggle dance, serves rather like the angular deviation of a celestial cue relative to the horizon.

The hive and combs are part of the system of computation required for effective foraging. And like an abacus and a slide rule, employ persistent features of physical geometry to allow for very precise behavior. A foraging bee is part body, part collective, and part physical hive. The functional unit of navigation -- analog to Chopin's Bauza preludes and Jarrett's Bosendorfer Koln concert –- is a behavior embedded in a life-constructed physics.

\section{Extending into Exbodiment} 

Ecologists study the interaction of species with their environments. Following Thomas Malthus, Charles Darwin had made environmental resource limitation the foundation of his regulatory theory of natural selection. Competition and cooperation in the face of environmental scarcity makes for adaptation. And the environment changes slowly against which organisms evolve relatively quickly.
Early formal ecologists realized that natural selection implies that the precise structure of the environment matters and that it should be described as carefully as the organism. Whereas organisms possess features such as anatomy and physiology, environments possess niches and trophic interactions - energy flows. 

\subsection{On the Origin of Niches}

Joseph Grinnel described the niche as an abiotic (non-living) recess in which the organism is tucked. Charles Elton sought to find some space in this recess for other organisms. But it was George Evelyn Hutchinson who made the first step toward formalizing the niche when he conceived of an "n-dimensional hypervolume" of survival-relevant resources. For Hutchinson, life might take place in three dimensions of space, but it is involved in vastly more dimensions of interaction and dependency: the dimensions of sunlight; water; oxygenation; etc. Hutchinson's dimensions are quantities analogous to the coordinates of space. They are the coordinates of existence. \cite{Leibold1995-uh}

It did not take too long for the question to arise, where do all these the niches come from? Grinnel had suggested that they come from physics. Elton and Hutchinson further proposed they also derive from other organisms. Richard Dawkins in 1978 \cite{Dawkins2016-wy}, Richard Lewontin in 1982 \cite{Lewontin2000-zl}, Hastings and colleagues in 1994 \cite{Hastings2007-dw}, and John Odling Smee and colleagues 1996 \cite{John_Odling-Smee2003-uk}, participated in a movement to undermine the whole organism environment dichotomy. According to these researchers, many environments are extended phenotypes or constructed niches or engineered ecosystems. The world is like a body that develops and grows out of both self and non-self. The niche had been a convenient fiction that makes natural selection efficient. The non-fiction account makes the niche an adaptive extension of the organism evolved to overcome the limitations of body and generate a super-physical system. In this telling the principle of selection is principle of exbodiment. 

\subsection{Spider Power}

In 2019 it was discovered that the web of the triangle weaver spider provides a potent source of external power amplification \cite{Han2019-pz}. The key idea is to slowly accumulate potential energy and then rapidly dissipate kinetic energy that can exceed the limits of physiology. The spider can increase the tension in the web silk by ratcheting their legs inwards through a leg-over leg loading motion. This iterative process takes place over prolonged intervals of time taking minutes to hours. When potential prey is caught in the web, this tension is released by rapid withdrawal of the limbs, causing the web to contract locally around the prey from all directions in a second or less. The spider-web extended system exploits the tensility and ductility of silk to overcome the kinematic constraints associated with hydraulic pressure exoskeletons. 

This is not an example of embodiment or embodied cognition which seeks to reduce cognitive overhead by outsourcing processing to the body, but exbodiment which expands physical and cognitive capability through the extension of the biological self into the shared and collectively intelligent physical world.

\subsection{The Meaning of External}
Niche construction and its cognates raises the obvious question, how are we to distinguish between organism and organism-built environment? In a paper from 2009 my colleagues and I answered this question with a mathematical theory of niche monopolies \cite{Krakauer2009-tb}.  A multicellular organism is comprised of reproductive cells - germ line - and tissue cells – soma. The key property of germ line is that it transmits heritable information across generations. Soma by contrast dies with the organism. Germ lines construct soma during development and hence soma is like any other material in the environment including a nest or a hive. With one key difference. By virtue of the spatial collocation of germ and soma cells in a body, all resources acquired by somatic tissue can be fed-back to the germ line with minimal loss. The same is not true of a nest or hive which can be appropriated by competitors. We desmonstrate that exbodiment can only evolve through niche construction if an organism can monopolize its niche the way a germ cells monopolizes resource-harvesting by soma. A body is nothing other than an exbodied artifact with a guaranteed closed feedback loop. 

\section{The Cortical Chopstick} 

Spiders and their webs, like birds and their nests, and bees and their hives, have evolved elaborate extended phenotypes which go some way to dissolving the boundary of organism and environment. The skills required to manipulate their exbodiment might be described as innate.

Humans on the other hand need to learn to use the tools they build. Chopin did not instinctively exude his Bauzer piano post-partum and then play it as naturally as wiggle his fingers. The cognitive and neural implications of this fact are intriguing. The ability to construct a physical artifact are largely dissociated mentally from its applications. This means that the neural encoding of a given "exbody" affordance needs to approximate the skill observed in typical internal adaptations, for example, the encoding of sound in the inner year, or smell in the olfactory bulb, or scenes in the visual cortex. This goes beyond constructed niches which modulate local physics, toward modifying through representation, the whole adaptive plan of the organism.

What might we expect to happen as we develop expertise with a common tool? Does the tool slowly morph into an exbodiment - an extension and amplification of limbs - and if so - how does the brain encode such a process? Sawamura and colleagues have sought to answer this question by tracking the acquisition of chopstick expertise is subjects forced to use their non-dominant hand \cite{Sawamura2021-rp}.  As trial subjects practise over a six-week period, concomitant with skill acquistion brain activity shifts from prefrontal cortex to premotor cortex. A pattern that is often interpreted as a shift from intentional and analytical solutions to motor programming more typical of limb motion. Expertise on this task elides manipulated tools and parts of the body. 

An earlier paper by Kitamura and colleagues on virtual chopsticks -- more readily manipulated into unconventional motion -- reached  similar conclusions \cite{Kitamura1999-kc} \cite{Kitamura2003-cy} Which suggests that our understanding of physical objects can also be understood in terms of kinematics - or spatial degrees of freedom - that live outside the confines of the body - physical or otherwise.

\section{Materializing the Tape} 

The first tape machines were invented in the 1880s with tapes made with wax paper. The wax was purloined from the hexagonal, navigationally inclined cones of beehives. Magnetic tapes with recording heads were produced shortly afterwards in the 1890s and moved into mass circulation in the 1930s. In both cases, signals are recorded by modifying the arrangement of a physical substrate on a surface for subsequent playback \cite{Fabrizio2005-iz}.

The first recordings of Chopin's piano music can be traced to cylindrical phonograph recordings made in 1919 by Alfred Cortot \cite{Nichols1982-fp}. Most likely recorded on a Steinway piano whose bombastic tones could not be further from the harp-like innocence of the Bauza.

Computer memory did not make use of ferric oxide audio tape until its use in the UNIVAC computer in 1951. Prior to UNIVAC computer memory was non-persistent and stored either in arrays of vacuum tubes or in the traveling waves of mercury delay lines \cite{McSkimin1948-ut}.

It is fair to assume that in 1936 when Alan Turing wrote his classic paper "On Computable Numbers, with an Application to the Entscheidungsproblem", that is with application to the decision problem, Turing had never used a tape recorder and most certainly not a computer. (Turing and Others 1936) The first university computer was the ENIAC completed at the university of Pennsylvania in 1945 \cite{Mauchly1980-tf}.

Turing's 1936 paper is a pessimistic answer to a question posed by the optimistic mathematician David Hilbert in 1928. Hilbert asked, and firmly believed any answer to be affirmative, whether there was a universal method, or mechanical process, that might demonstrate that any mathematical assertion is true or false. Turing showed that this question was in the most general sense unanswerable, that is undecidable.

The way that Turing solved the Hilbert problem is beautifully physical and not at all like the abstractions that we typically associate with mathematical proof. In order to demonstrate undecidability Turing invented a fictitious computing machine that calculates by modifying persistent marks on a tape -- Turing internalized exbodiment.  With his rare intelligence Turing uses a physically impossible infinite model inspired by a finite physical device to refute an infinite proposition.

The key elements of the universal Turing machine are the infinite tape which includes its programs, the data upon which it will operate; a tape head that reads and writes to the tape according to its program, and a halting criterion which ends the programs with a true or false statement. Turing proved that there are problems that can be presented in the data for which the machine will never halt. These problems are formally undecidable - they have no computable solutions.

Unlike the web and the chopstick that through evolution or learning could become an exbodied part of a spider or a human, both of which extend beyond the limits of the physical, the Turing tape is a reabsorbed physical device. And by virtue of its abstraction, it can be extended beyond physical limits into the spooky expanse of the infinite. The Turing Tape is the perplexing imagination of an mental exbodiment. And it is one that serves to illustrate both the flexibility and limits of our reason.

\section{Alien Artifacts} 

One of the more obvious features of human activity has been the construction of very large-scale material and inferential structures - to include, powerplants, cities, the internet, the world wide web, and stored knowledge in the form of text and electromagnetic digital media. Some have even argued that these kinds of activities represent generic techno signatures providing information useful for the detection of intelligent life in the universe \cite{Benford2021-vh} \cite{Lingam2021-jy}.

An intriguing dilemma related to the use of electromagnetic techno signatures, is as some of my colleagues have proved, that efficiently encoded signals tend to be indistinguishable from noise \cite{Lachmann2004-jz}. The efficiency of information-coding tends to minimize redundancies that might provide clues that a signal is of non-astrophysical origin. Furthermore, a problem confronting all non-physical signs of life, is how we should conclusively rule out rare physical events mimicking a life-like syntax?

\subsection{Irreducible Physicality}

One way around this problem is to search for evidence of unlikely physical structures, to include macro-engineering and megastructures, such as the Stapledon-Dyson sphere or starshades \cite{Lingam2018-jz}.  Perhaps more interesting is to search for objects of sufficient complexity that we might confidently rule out natural and non-purposeful physical origination. For example, if we found a piano on the surface of Mars would that not be conclusive? As Charles Bennet has suggested, even finding a safety pin on the surface of a planet would be conclusive. This is because, as Adam Smith observed for the manufacture of anything efficiently, the division of labour and use of design is often bewildering \cite{Smith2019-km}:

\begin{quote}
    One man draws out the wire, another straights it, a third cuts it, a fourth points it, a fifth grinds it at the top for receiving the head; to make the head requires two or three distinct operations; to put it on, is a peculiar business, to whiten the pins is another; it is even a trade by itself to put them into the paper; and the important business of making a pin is, in this manner, divided into about eighteen distinct operations, which, in some manufactories, are all performed by distinct hands’" – Adam Smith.
\end{quote}

James Benford has recently even called for a new discipline of ET archaeology \cite{Benford2021-vh}. We should be studying high resolution images of the moon, and nearby orbiting objects, for signs of past and present intelligent life. On the grounds that a physical fossil will always be more conclusive than a signal. The same objective underlies recent efforts to quantify time through spatial extensivity in materials as gauged by the Assembly Index \cite{Marshall2022-sx}. 

\section{The Helix of Exbodiment } 

\includegraphics[scale=0.3]{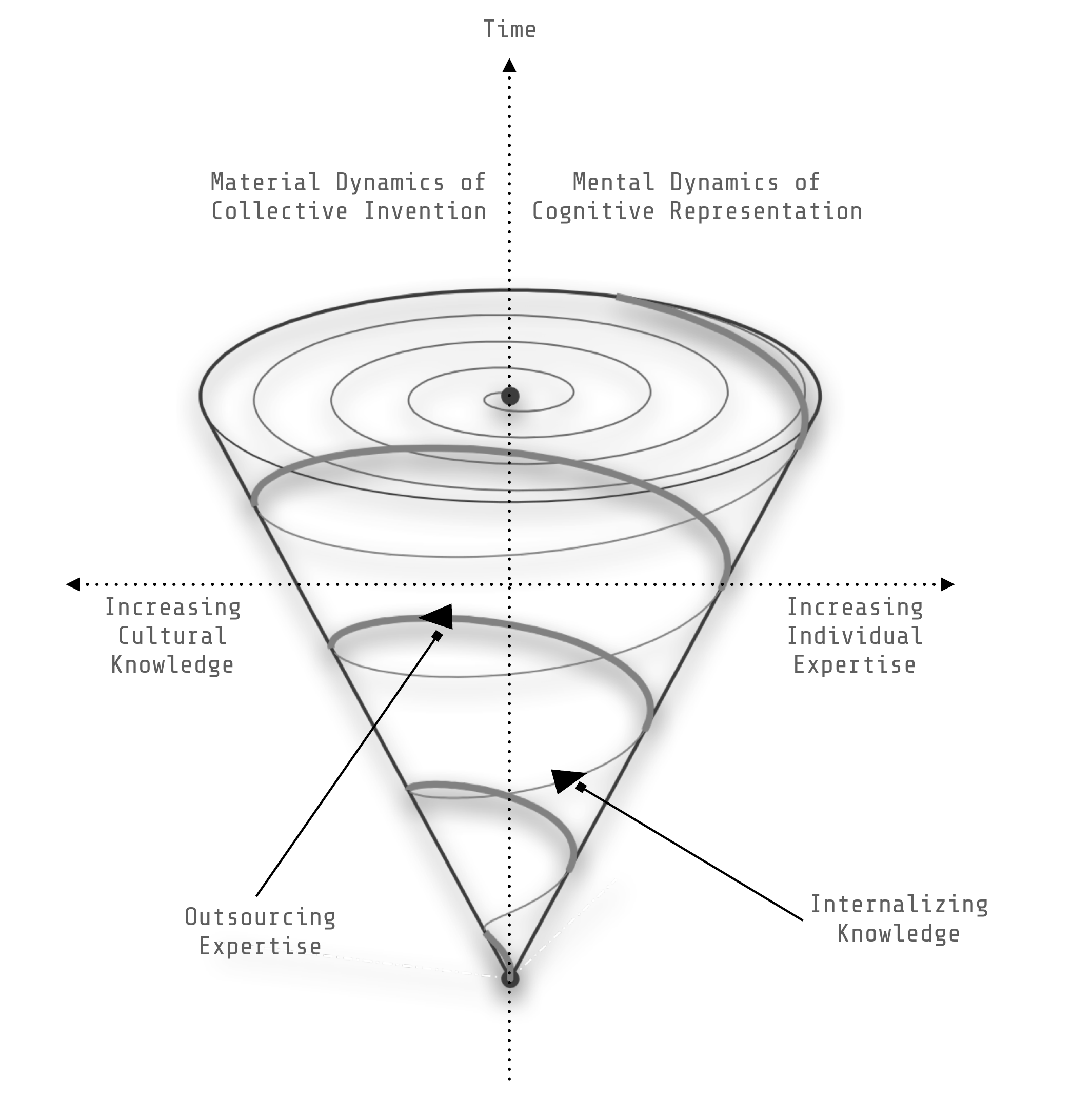}

The interaction between the interior mind and the exterior engineered artifact is ongoing: improvements in mind lead to modifications of design, whereas improvements in artifact lead to increases in the acquisition of skill and expertise. This feedback might be visualized as a helix bisecting the exterior "material dynamics of collective invention" (artifacts are a part of culture) and the interior "mental dynamics of cognitive representation" (minds are carried by individuals). In time both domains are expanded through practice. 

Acquiring expertise in using the abacus and playing chess  illustrate the helix of exbodiment. Acquiring abacus skills requires use of a physical abacus but continued use leads to the formation of a "mental abacus" that can be accessed without an artifact \cite{Hanakawa2000-at}. Expert users consolidate a variety of distinct circuits for complicated tasks versus simple ones whereas novices make use of a single circuit \cite{Hanakawa2003-pf}. Moreover the mental abacus facilitates a range of non-arithmetic cognitive tasks \cite{Chen2006-ba}. Similarly Chess is first learned with a board, through practice can be played blindfolded \cite{Fine1965-eu}, and with continued board-based play leads to higher resolution encoding for strategically significant configurations on a purely mental board \cite{Saariluoma1998-pz}.

\section{Self Construction}

In his novel Solaris \cite{Lem2016-jk}, Stanislav Lem captured many of the challenges that connect materiality with the pursuit of problem solving intelligence. Concluding that it is not physical special effects that connote intelligence but material self-doubt. Through an apparently endless Helix of Exbodiment the planet Solaris mentates and manifests at once. 

\begin{quote}
   For some time, there was a widely held notion (zealously fostered by the daily press) to the effect that the ‘thinking ocean’ of Solaris was a gigantic brain, prodigiously well- developed and several million years in advance of our own civilization, a sort of ‘cosmic yogi,’ a sage, a symbol of omniscience, which had long ago understood the vanity of all action and for this reason had retreated into an unbreakable silence. The notion was incorrect, for the living ocean was active. Not, it is true, according to human ideas — it did not build cities or bridges, nor did it manufacture flying machines. It did not try to reduce distances, nor was it concerned with the conquest of Space (the ultimate criterion, some people thought, of man’s superiority). But it was engaged in a never-ending process of transformation, an ‘ontological autometamorphosis.’ 
\end{quote}

\section{Acknowledgements. }
This work is in part supported by the Templeton World Charity Foundation, Inc. (funder DOI 501100011730) under the grant https://doi.org/10.54224/20650 and no. 20650 on Building Diverse Intelligences through Compositionality and Mechanism Design, and
grant no. 81366 from the Robert Wood Johnson Foundation on Using Emergent Engineering for integrating complex systems to achieve an
equitable society.

\end{document}